# Co-occurring Diseases Heavily Influence the Performance of Weakly Supervised Learning Models for Classification of Chest CT

Fakrul Islam Tushar[a,b], Vincent M. D'Anniballe[a],
Geoffrey D. Rubin[c], Ehsan Samei[a,b], Joseph Y. Lo[a,b]

[a]Center for Virtual Imaging Trials, Carl E. Ravin Advanced Imaging Laboratories,
Department of Radiology, Duke University School of Medicine, Durham, NC.
[b]Department of Electrical and Computer Engineering, Pratt School of Engineering, Duke University, Durham, NC
[c]Department of Medical Imaging, University of Arizona, Tucson AZ.

**Abstract**

Despite the potential of weakly supervised learning to automatically annotate massive amounts of data, little is known about its limitations for use in computer-aided diagnosis (CAD). For CT specifically, interpreting the performance of CAD algorithms can be challenging given the large number of co-occurring diseases. This paper examines the effect of co-occurring diseases when training classification models by weakly supervised learning, specifically by comparing multi-label and multiple binary classifiers using the same training data. Our results demonstrated that the binary model outperformed the multi-label classification in every disease category in terms of AUC. However, this performance was heavily influenced by co-occurring diseases in the binary model, suggesting it did not always learn the correct appearance of the specific disease. For example, binary classification of lung nodules resulted in an AUC of < 0.65 when there were no other co-occurring diseases, but when lung nodules co-occurred with emphysema, the performance reached AUC > 0.80. We hope this paper revealed the complexity of interpreting disease classification performance in weakly supervised models and will encourage researchers to examine the effect of co-occurring diseases on classification performance in the future.

**Keywords:** co-occurring diseases, multi-label classifier, binary classifier, weak-supervision, classification, CT.

## Introduction:

Despite the high performance of artificial intelligence (AI) models in medical imaging, most models remain narrow in scope, with the majority targeting one disease or organ, ignoring the correlations and co-occurrence of the diseases. In response to these limitations, recent computer-aided diagnosis models have leveraged weak, noisy text annotations extracted from radiology reports [1-6]. In our earlier study, we were one of the first to classify multiple diseases in lungs/pleura, liver/gallbladder, and kidneys/ureter using weakly supervised classification labels extracted using rule-based algorithms (RBA) [5, 6]. The main contribution of this paper is that we have investigated differences between binary and multi-label classification on a multi-label dataset by measuring the influence of disease co-occurrence on binary classification.

## Methodology:

### Label Extraction, Dataset, and Preprocessing:

In this study, we focused on a multi-label dataset of the lungs/pleura outlined in our previous study [5]. Tushar *et al*. using RBA, extracted 7,441 disease labels from the text reports of 5,044 body CT scans of 4,639 unique subjects [5, 6]. The rules of the RBA were built upon general relationships, which can be applied to different organs by adding a dictionary structure of organ-specific keywords. For example, if a sentence contained the organ descriptor "lung" or

"lobe" and abnormality "nodule" and there was no negative like "no" or "without", then the label was positive for lung nodule. The lungs/pleura were labeled as no apparent disease or having one or more of four diseases: atelectasis, nodule, emphysema, and/or effusion. The RBA labeled a report as no apparent disease only in the absence of dozens of other diseases that were not otherwise analyzed used in this study. The total number of volumes associated with each disease and normal label used in classification tasks were as follows: 1194 for emphysema, 1465 for effusion, 1628 for nodule, 1758 for atelectasis and 1396 for no apparent disease. **Fig. 1** illustrates the co-occurrence and association between diseases for this multi-label data set.

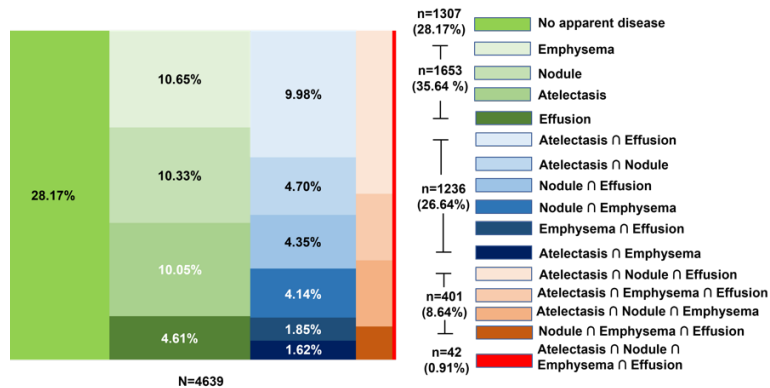

**Figure 1**. The tree diagram displays the occurrence and co-occurrence of targeted diseases among unique subjects in the lungs/pleura. The area corresponds to the frequency for each combination of diseases. N= number of unique subjects per organ, n= number of unique subjects belonging to that (co-)occurrence. The percentage of n/N is noted in parentheses inside each box.

Prior to experimentation, all CT volumes were resampled to voxels of size 2 *mm* x 2 *mm* x 2 *mm* via B-spline interpolations, clipped to intensity range (-1000, 800) HU, and normalized to 0 mean and 1 standard deviation. The CT volumes were randomly divided into subsets to train (70%), validate (15%), and test (15%) the model. Splitting was performed by subject and separately for normal vs. diseased classes to preserve disease prevalence across each subset.

**Classification workflow and 3D CNN:**

Fig. 2 illustrates the overall classification workflow. First, applying an RBA [6] to each radiology report associated with a CT scan allowed disease label acquisition. Afterward, a segmentation module was trained in a semi-supervised manner [5], which provided the lungs segmentation mask for guided patch extraction for classification. The classification CNN used in this study is a 4-resolution 3D CNN inspired by ResNet.

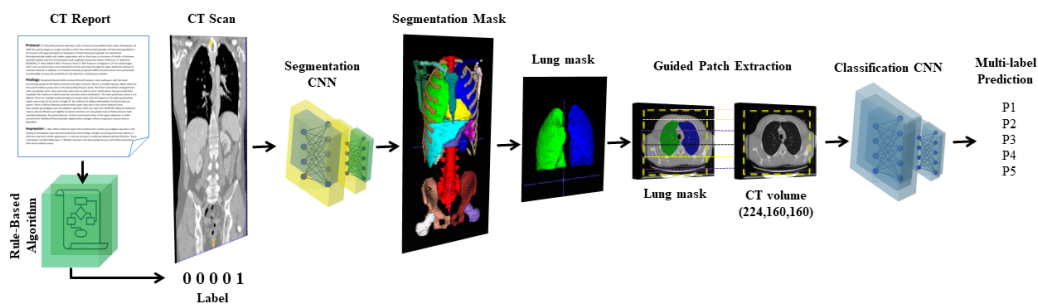

**Figure 2**. Study overview. Illustration of weakly supervised classification study workflow. 4 abnormalities and no apparent disease labels were extracted using rule-based algorithms. A segmentation model was used to extract

lung patches from body CTs. Finally, we trained and validated 3D CNN for classification on 3D CT volumes to classify all four abnormalities and no apparent disease simultaneously.

**Binary and Multi-label experiments:**

Our dataset is characterized by high co-occurrence and association between diseases as shown in fig. 1, which is natural in medical scenarios. With our dataset, we have compared multi-label classification (MLCL) and binary classification (BCL) and performed a series of experiments to understand and analyze the effect of co-occurrence on performance. Table 1 shown the classifications performed and classification definitions.

**Table 1:** Classification models annotations and classification task definition.

| Classification Model | Classification task |
|---|---|
| Multi-label classification (MLCL) | Model simultaneously classify all four disease classes: atelectasis, nodule, effusion, emphysema, and no apparent disease class. |
| Binary classification (BCL) | Model classify no apparent disease vs. abnormalities combining disease classes atelectasis, nodule, emphysema, and effusion |
| Binary nod. classification (BNCL) | Model classify nodule (All nodule positive) Vs. no apparent disease. |
| Binary nod. vs. non-nod. classification (BNNCL) | Model classify nodule (All nodule positive) vs. non-nodule (combined atelectasis, emphysema, effusion, and no apparent disease) |

## Results:

The test set consists of 771 CT volumes with 1154 labels. All 1,154 labels in the test set were manually validated. The number of positive examples for each class were as follows: 251 for atelectasis, 296 for nodule, 193 for Emphysema, 205 for Effusion, and 209 for no apparent disease.

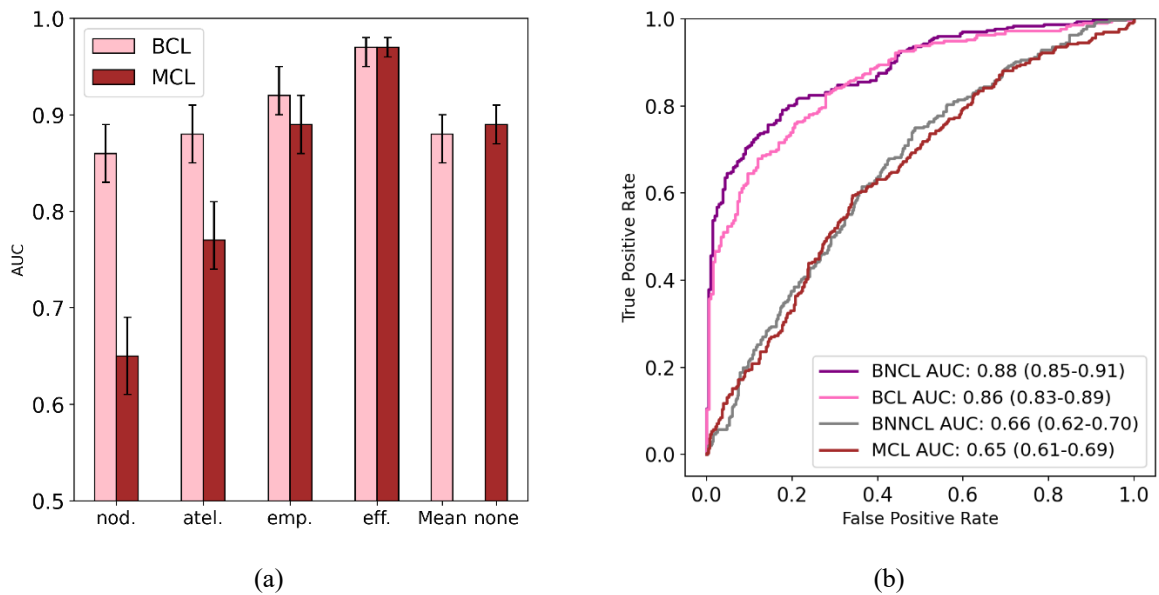

**Figure 3. (a)** Classification performance on test set of binary and multi-label classification. Error Bar represents the 95% confidence interval. Note that binary (pink) is always greater than or equal to multi-label (brown) performance.

**(b)** Classification performance of Nodule Class with different experiments. Nodule, atelectasis, emphysema, effusion, and no apparent disease were denoted as nod., atel., emp., eff. and none. respectively.

Fig. 3 shows the performance of the binary and multi-label classifications. For emphysema and effusion, the binary and multi-label model performed similarly. In contrast, there is a significant performance increase in binary classification compared to multi-label for nodule and atelectasis.

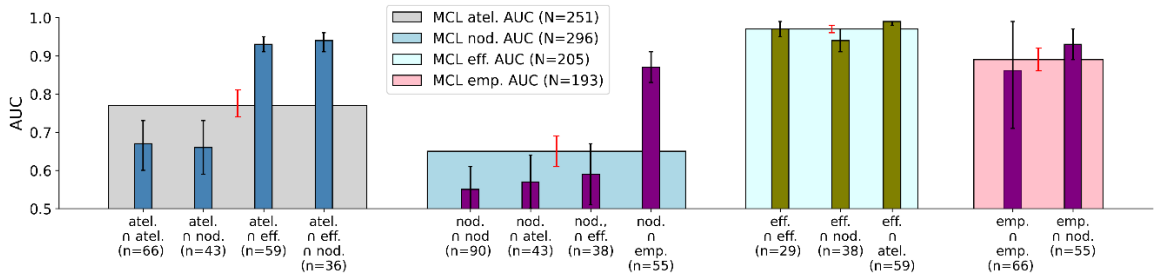

**Figure 4.** Performance of the multi-label classification (MCL) model on the test set based on the single, two, and three class significant associations. **N**=number of total positive volumes associated with that class, **n**=numbers of positive volume related to the association; error bar represents 95% confidence interval. Nodule, atelectasis, emphysema, and effusion were denoted as nod., atel., emp., and eff. respectively.

Fig. 4 shows the performance of the disease classes in terms of exclusivity and co-occurrence with other classes. When the nodule class co-occurs with emphysema, it always results in an increase in performance > 0.80, whereas with other classes its performance is marked worse at < 0.60. Similar relationships can be observed between atelectasis and effusion, where atelectasis performance is higher when co-occurring with effusion.

### Discussion and future work:

In this work, we have investigated the differences between binary and multi-label classifiers. In this comparison, the binary classifier outperformed the multi-label classification for focal disease classes (e.g., nodule). As measured by AUC, performance does not necessarily indicate that the multi-label disease classifier is worse. This is evidenced by the nature of our dataset, where the nodule class overlaps a lot with other diseases, and the binary model was able to pick those others easier, larger diseases, but that performance did not reflect actual classification of nodules per se. Another fascinating finding of our study, as illustrated in fig 4., is that the multi-label model effectively benefited from the clinically significant disease co-occurrences, such as the association of atelectasis with effusion, which always resulted in a better identification of atelectasis in terms of AUC. Similar trends can be seen for the emphysema and nodule class where the association of nodules with emphysema always results in a better nodule performance. Our future efforts will investigate indemnifying features introducing bias in model decision making, clinically relevant relationships between co-occurring abnormalities.

### Acknowledgment:

This work was funded in part by developmental funds of the Duke Cancer Institute as part of the NIH/NCI P30 CA014236 Cancer Center Support Grant, as well as the Center for Virtual Imaging Trials, NIH/NIBIB P41-EB028744.